\documentstyle[12pt]{article}
\def\be{\begin{equation}}       
\def\ee{\end{equation}}
\def\bear{\be\begin{array}}      
\def\eear{\end{array}\ee}
\def\bea{\begin{eqnarray}}
\def\eea{\end{eqnarray}}

\def\phii{\mbox{$\phi$}}

\def\n{\mbox{$N$}}

\def\npb#1#2#3{\mbox{Nucl. Phys. {\bf B#1} (#2) #3}}
\def\plb#1#2#3{\mbox{Phys. Lett. {\bf B#1} (#2) #3}}
\def\prd#1#2#3{\mbox{Phys. Rev. {\bf D#1} (#2) #3}}
\def\prl#1#2#3{\mbox{Phys. Rev. Lett. {\bf #1} (#2) #3}}
\def\prc#1#2#3{\mbox{Phys. Rep. {\bf #1} (#2) #3}}
\def\zpc#1#2#3{\mbox{Z. Phys. {\bf C#1} (#2) #3}}
\def\ijmp#1#2#3{\mbox{Int. J. Mod. Phys. {\bf A#1} (#2) #3}}

\begin{document}

\baselineskip 18pt

\newcommand{\sheptitle}
{\Large A Next-to-Minimal Supersymmetric
Model \\ of Hybrid Inflation ${}^{\dag}$}

\newcommand{\shepauthor}
{M. Bastero-Gil${}^1$ and S. F. King${}^{2*}$}

\newcommand{\shepaddress}
{${}^1$Department of Physics and Astronomy, University of Southampton\\
Southampton, SO17 1BJ, U.K. \\ 
${}^2$ CERN, Theory Division, CH--1211 Geneva, Switzerland}

\newcommand{\shepabstract}
{
In this talk we discuss
a model of inflation based on a simple variant of the NMSSM,
called $\phi$NMSSM \cite{phiNMSSM}, where the additional singlet $\phi$
plays the role of the inflaton in hybrid (or inverted hybrid) type models. 
As in the original NMSSM, the $\phi$NMSSM solves the
$\mu$ problem of the MSSM via the VEV of a gauge singlet $N$,
but unlike the NMSSM does not suffer from domain wall problems
since the offending $Z_3$ symmetry is replaced by an approximate
Peccei-Quinn symmetry which also solves the
strong CP problem, and leads to an invisible axion with 
interesting cosmological consequences.
The model predicts a spectral index $n=1$ to one part in $10^{12}$.}

\begin{titlepage}
\begin{flushright}
CERN-TH/98-25\\
hep-ph/9801451\\
\end{flushright}
\vspace{.1in}
\begin{center}
{\large{\bf \sheptitle}}
\bigskip \\ \bigskip\shepauthor\bigskip \\ 
\mbox{} \\ {\it \shepaddress} \\ \vspace{.5in}
{\bf Abstract} \bigskip \end{center} \setcounter{page}{0}
\shepabstract\\
\begin{flushleft}
CERN-TH/98-25\\
January 1998
\end{flushleft}

\noindent ${}^{\dag}$Talk given by S.F.K. at COSMO-97,
International Workshop on Particle Physics and the Early Universe,15-19
September 1997, Ambleside, Lake District, England.\\
\noindent${}^*$ On leave of absence from $^3$.

\end{titlepage}

\setcounter{page}{1}

%
\def\npb#1#2#3{\mbox{Nucl. Phys. {\bf B#1} (#2) #3}}
\def\plb#1#2#3{\mbox{Phys. Lett. {\bf B#1} (#2) #3}}
\def\prd#1#2#3{\mbox{Phys. Rev. {\bf D#1} (#2) #3}}
\def\prl#1#2#3{\mbox{Phys. Rev. Lett. {\bf #1} (#2) #3}}
\def\prc#1#2#3{\mbox{Phys. Rep. {\bf #1} (#2) #3}}
\def\zpc#1#2#3{\mbox{Z. Phys. {\bf C#1} (#2) #3}}
\def\ijmp#1#2#3{\mbox{Int. J. Mod. Phys. {\bf A#1} (#2) #3}}


There is to date no standard model of inflation, and although there has
been a good deal of progress in recent years in this area much of 
the current activity 
has been concerned with conceptualised field theoretic
models rather than well motivated particle physics based models
\cite{kolbturner}.
Possibly the best motivated particle physics model beyond the
standard model is the minimal supersymmetric standard model (MSSM).
However the only Higgs fields in the MSSM are the two doublets
$H_1,H_2$, which develop vacuum expectation values (VEVs) of order the
weak scale, and it is very difficult if not impossible to develop
a model of inflation using only these fields for several reasons.
The primary reasons are that the electroweak scale turns out to
be too small and the Higgs potential is not sufficiently flat.
The so called next-to-minimal supersymmetric standard model
(NMSSM) is more promising from the point of view of inflation
since it contains, in addition to the two Higgs doublets, a Higgs 
singlet $N$ which may develop a large VEV. 

The usual NMSSM does not require the 
$\mu H_1H_2$ term of the MSSM, replacing it with
a $\lambda NH_1H_2$ term, and thereby solving the
$\mu$ problem\footnote{Note that the Giudice-Masiero mechanism \cite{mu}
presents a solution to the
$\mu$ problem within the MSSM
by generating the $\mu$ term via a non-minimal
Kahler potential.}.  
The NMSSM also involves a
term $k N^3$ in the superpotential
so that the model has an exact $Z_3$ symmetry
\cite{NMSSM,NMSSMphenom}. However this is broken at the weak scale
leading to a serious domain wall problem \cite{walls,NMSSMwalls}.
Originally it was thought
that the $Z_3$ may be slightly violated by Planck scale operators,
leading to a pressure term that removes the walls. However without an exact
$Z_3$ symmetry supergravity tadpole diagrams will lead to a large
singlet mass in the low energy theory, and the amount of $Z_3$ breaking
required to solve the domain wall problem is in conflict with
requirement that tadpoles do not make the singlet too heavy 
\cite{NMSSMwalls,various}. 

It transpires that, without fine-tuning, the NMSSM does not lead
to a sufficiently flat potential along which the inflaton may roll.
In order to overcome this we introduce a second singlet $\phi$,
and replace the term $N^3$ in the NMSSM by $\phi N^2$.
Thus our model is based on the superpotential:
\begin{equation}
W_{\phi NMSSM} =\lambda NH_1H_2 -k \phi N^2
\label{phiNMSSM}
\end{equation}
Note that our model has the same number of dimensionless couplings as the
original NMSSM, and we have used the same notation $\lambda ,k$ to emphasise 
this. With this modification the
field $\phi$ appears only linearly in the
superpotential and so will have a very flat potential, lifted only by
a tiny mass $m_{\phi}$ of order electronvolts, and will play the role
of the inflaton field of hybrid inflation 
\cite{hybrid,copeland,morehybrid} if 
$m_{\phi}^2>0$ or inverted hybrid inflation \cite{inverted} if
$m_{\phi}^2<0$. In the case of inverted hybrid inflation the present
model provides an interesting counter example to the problems
raised in Ref. \cite{invertedproblems}.
Inflation ends when $\phi$ reaches a critical value
$\phi_c \sim 10^{13}$ GeV after which the $N$ field, which has a zero
value during inflation, develops a VEV $<N> \sim \phi_c$.
Interestingly the inflaton also develops an eventual VEV
$<\phi > \sim \phi_c$ via a tadpole coupling, which is 
typical of inverted hybrid inflation but quite extraordinary for
hybrid inflation.The resulting dimensionless couplings are
$\lambda , k \sim 10^{-10}$, whose smallness will be explained
by embedding the model into a string inspired model where the couplings 
result from higher dimension operators, 
controlled by discrete symmetries \cite{phiNMSSM}.
Note that radiative corrections to the inflaton mass are controlled
by $\lambda , k$ and are of order the inflaton mass itself.

Having replaced the NMSSM superpotential by 
Eq. (\ref{phiNMSSM}), the troublesome $Z_3$ symmetry 
is replaced by a global $U(1)_{PQ}$ Peccei-Quinn symmetry
where the global charges of the fields satisfy:
\begin{equation}
Q_N+Q_{H_1}+Q_{H_2}=0, \ \ Q_{\phi}+2Q_N=0.
\label{PQcharges}
\end{equation}
with the quark fields having the usual axial PQ charges.
The global symmetry forbids additional couplings such as $N^3$, $\phi H_1H_2$
and so on, but is broken at the scale of the VEVs releasing a very light
axion. The axion scale $f_a$ is therefore of order $\phi_c$ in this model.
The axion will be an invisible 
Dine-Fischler-Srednicki-Zhitnitskii (DFSZ) \cite{dfsz} type axion,
which couples to ordinary matter through its mixing with the standard
Higgses after the electroweak phase transition.
Once we embed our model into a string motivated model, the global
PQ symmetry will emerge as an approximate accidental symmetry of an underlying
discrete symmetry, and we need to discuss such questions as the solution
to the strong CP problem in this wider context.
Note that if we had simply 
removed the $N^3$ term from the NMSSM superpotential
and not replaced it with anything
then the theory would also have a PQ symmetry,
and the potential would also be flat in the $N$ direction, 
and then one might be tempted to identify $N$ with the
inflaton of hybrid inflation. However in such a scenario the height of the
potential during inflation would be of order 1 TeV, leading to
an inflaton mass very much smaller than the radiative corrections
to its mass of order 1 eV, which would require unnatural fine-tuning. 
By contrast, with the $\phi N^2$ term present,
the height of the potential during inflation is about $10^8$ GeV
and the COBE constraint may be satisfied by an inflaton mass of about 1 eV
which is the same order as the radiative corrections to its mass,
leading to a natural scenario with no fine-tuning required,
as discussed in \cite{phiNMSSM}.

The tree-level potential which follows from the superpotential in 
Eq. (\ref{phiNMSSM}) can be written, 
if we ignore $H_1,H_2$ which have smaller VEVs,
\begin{eqnarray}
V_0 & = & V(0)+V(\phi ,N) \nonumber \\
  V(\phi,N)&=& k^2 N^4 + m^2(\phi) N^2 +m_{\phi}^2\phi^2
\,, \label{vphin2}
\end{eqnarray}
with the field dependent $N$ mass given by,
\begin{equation}
  m^2(\phi)= m_N^2- 2 k A_k \phi + 4 k^2 \phi^2 \,.
\end{equation}
We have taken \phii\ and \n\ to be the real components of the
complex singlets, and included the soft breaking parameters from the
soft supersymmetry breaking potential terms
$m_NN^2$, $m_{\phi}\phi^2$ and $A_kk\phi N^2$.
We have also added by hand a constant vacuum energy $V(0)$ to the potential,
about which we shall say more later.
Note that $m^2(\phi)=0$ for
\phii\ equal to a critical value\footnote{We require that the condition
$A_k^2 > 4 m_N^2$ is fulfilled.}: 
\begin{equation}
 \phi_c^{\pm}=\frac{A_k}{4 k} \left( 1 \pm \sqrt{1-4 \frac{m_N^2}{A_k^2}}
 \right) \,.
\label{phic}
\end{equation}

In order to discuss inflation we need to specify the sign of
the inflaton mass squared $m_{\phi}^2$. If $m_{\phi}^2>0$ 
(as in hybrid inflation) then,
for $\phi > \phi_c^+$, \n\ will be driven to a local
minimum (false vacuum) with \n=0. 
Having a positive mass squared, $\phi$ will roll towards the origin
and $m^2(\phi)$ will become
negative once the field \phii\ reaches $\phi_c^+$.
After that, the potential develops an instability in the \n=0
direction, and both singlets roll down towards the global minimum, 
\begin{eqnarray}
   <\phi>&=& \frac{A_k}{4 k} \,,\\
    <N>  &=& \frac{A_k}{2\sqrt{2} k}\sqrt{1-4 \frac{m_N^2}{A_k^2}}
=\sqrt{2} \left|
   \phi_c^{\pm}-<\phi > \right|\,,
\end{eqnarray}
signaling the end of the inflation. 
On the other hand if $m_{\phi}^2<0$ 
(corresponding to inverted hybrid inflation)
then we shall suppose that 
during inflation $\phi < \phi_c^-$, with the inflaton rolling away from the
origin, eventually reaching $\phi_c^-$ and ending inflation
with the same global minimum as before. Note that the global minimum
VEV $<\phi>$ is sandwiched in between $\phi_c^-$ and $\phi_c^+$
so either hybrid or inverted hybrid inflation is possible in this
model depending on the sign of $m_{\phi}^2$.

Since $A_k$ is a soft SUSY breaking parameter of order 1 TeV we have the
order of magnitude results:
\begin{equation}
k\phi_c^{\pm} \sim k<N> \sim k<\phi > \sim 1 \ TeV.
\end{equation}
Since the VEVs are associated with the large axion scale, we see that the
parameter $k\sim O(10^{-10})$. Similarly since $\lambda <N>$ plays the
role of the $\mu$ parameter of the MSSM we require
$\lambda $ to have a similarly small value. 
We shall discuss the origin of such a small
values of $\lambda, k$ later in the context of the string motivated model, 
but for now we simply note their smallness and continue.

The negative value of
$V(\phi,N)$ at the global minimum, 
is compensated by $V(0)$ which is assumed to take an equal and opposite
value, in accordance with the observed small cosmological constant.
Thus we assume:
\begin{equation}
V(0)=- V(<\phi >,<N>)= k^2<N>^4=4 k^2 (\phi_c^{\pm} - <\phi >)^4 .
\label{V0}
\end{equation}
During inflation we may set the field 
$N=0$ so that the potential simplifies to:
\begin{equation}
 V= V(0) + m^2_\phi \phi^2
\end{equation}
The slow roll conditions are given by:
\begin{equation}
\epsilon_N=\frac{1}{16\pi}\frac{M_P^2{m_\phi}^4\phi_N^2}{V(0)^2}\ll 1,
\label{epsilon}
\end{equation} 
\begin{equation}
|\eta_N | =\frac{M_P^2}{8\pi} \frac{|m_{\phi}^2|}{V(0)} \ll 1.
\label{eta}
\end{equation}
The subscripts ``N'' means 
that $\phi$ and $\epsilon$ have to be evaluated  N e-folds
before the end of inflation, when the largest scale of cosmological
interest crosses the horizon that is, $N\simeq 60$. 
The height of the potential during inflation
is approximately constant and given by
$V(0)^{\frac{1}{4}}=k^{\frac{1}{2}}<N> \sim 10^8$ GeV.

Assuming that
$V(0)$ dominates the potential during inflation,
$\phi_N = \phi_c^{\pm} e^{\eta N}$ hence
$\phi_N \approx \phi_c^{\pm}$,
since in our model $|\eta | \ll 1/N$.
We need further to check that our inflationary model is able to
produce the correct level of density perturbation, responsible for the
large scale structure in the Universe, accordingly to the COBE
anisotropy measurements. 
The spectrum of the density perturbations is given by the quantity
\cite{deltah}, 
\begin{equation}
\delta_H^2= \frac{32}{75}\frac{V(0)}{M^4_P}\frac{1}{\epsilon_{N}}\,,
\end{equation}
with the COBE value, $\delta_H= 1.95 \times 10^{-5}$ \cite{cobe}. 
Writing $\phi_c^{\pm}\sim \phi_c$,
COBE gives the order of magnitude constraint:
\begin{equation}
 | k m_\phi | \simeq 8 \left(\frac{8 \pi}{75}\right)^{1/4} \delta_H^{-1/2}
  \frac{ (k \phi_c)^{5/2}}{M_P^{3/2}} \simeq 10^{-18} \ GeV \ 
  \left( \frac{k \phi_c}{1\, TeV} \right)^{5/2}\,.
\end{equation}
This, in turn, is more than enough to broadly satisfy the slow-roll
conditions. In particular, 
\begin{eqnarray}
|\eta_N | & 
\simeq &\frac{M_P^2}{8 \pi} \frac{|k m_\phi|^2}{(\sqrt{2} k \phi_c)^4}
\sim 10^{-12} \,, \\
\epsilon_N & \sim & \frac{M_p^2}{16 \pi} \frac{|k m_\phi|^4}{(\sqrt{2}k
\phi_c)^8} 
\phi_N^2 \sim 4 \pi \frac{ \phi_N^2}{M_P^2} \eta^2_N
\end{eqnarray}
The model predicts a very flat spectrum of density perturbations, as
usual in this type of hybrid model, with no appreciable deviation of
the spectral index, $n=1 + 2 \eta - 6 \epsilon$, from unity. Only
models where the curvature (of either sign) of the inflaton potential
is not very suppressed with respect to $H$ can give rise to a blue
\cite{bellido} (red \cite{driotto}) tilted spectrum.

Note that COBE requires the product $|km_\phi |$ to be extremely small.
If we take $k\sim 10^{-10}$, motivated by axion physics as discussed above,
then this implies $m_{\phi}$ in the electronvolt range. 
The requirement of such a small mass leads to several interesting
requirements on the model. We envisage that at the Planck scale
the $\phi$ mass is equal to zero. This can be
naturally accomplished within the framework of supergravity
no-scale models 
\cite{noscale}, where $some$  (not necessarily all) of the SUSY soft
masses are predicted to vanish, but with non-zero and universal
trilinear coupling parameters. The high energy value of 
$m_{\phi}$ will be subject to radiative corrections which are
very small, being controlled by the small coupling $k$,
leading to a mass $m_{\phi}$ in the eV range \cite{phiNMSSM}
for $k \approx 10 ^{-10}$. The small coupling leads to
a low reheating temperature $T_{RH}\approx O(1-10)$ GeV \cite{phiNMSSM}.
Despite its low value, the reheat temperature
is  high enough to preserve the standard scenario for
nucleosynthesis, $T_{RH} > $  6 MeV, although quite far to allow electroweak
baryogenesis. Moreover, any pre-existing baryon asymmetry is likely to
be diluted during inflation. 
Nevertheless, as has been pointed out \cite{kim,driotto}, the amount of
baryon asymmetry needed might be produced 
directly by the decays of the inflaton. For this
mechanism to work we require the presence of baryon-number violating
operator in the superpotential, type $\lambda_{ijk}'' U_i^c D_j^c
D_k^c$. As discussed the inflaton can decay predominantly
into light stop squarks, and the subsequent decay of the stops into
two down-type quarks from this R-parity baryon number violating
operator will generate baryon-antibaryon asymmetry.
Other mechanisms, like Affleck-Dine type baryogenesis
\cite{affleckdine}, might also work.  

Finally we summarise the successes and open problems
facing the model, and indicate some promising new directions 
along which progress may be made. The model in Eq.\ref{phiNMSSM}
represents a simple variant of the NMSSM and has the same number
of dimensionless coupling constants. However, unlike the NMSSM,
it does not have a domain wall problem since the discrete $Z_3$
symmetry has been replaced by a continuous PQ symmetry, thereby
solving the strong CP problem at the expense of raising the singlet
VEVs to $10^{13}$ GeV, and tuning down the dimensionless
couplings to $10^{-10}$. We have shown that with these parameters,
plus a very light $\phi$ mass in the eV range, a satisfactory
model of hybrid (or inverted hybrid)
inflation may result with the prediction that the spectral index $n$
is indistinguishable from unity. The model immediately raises a number
of questions, some of which were answered in the fuller treatment
in ref.\cite{phiNMSSM}, and some which remain open problems.
One question which has been addressed is that of the
origin of the small couplings $\lambda$ and $k$, whose smallness can
be understood as the result of certain (string inspired)
discrete and gauge symmetries which forbid the
operators $NH_1H_2$ and $\phi N^2$ at the renormalisable level,
but which allow similar non-renormalisable operators involving
additional singlets. The smallness of the ratio of the VEVs of the new 
singlets to the Planck scale then explains the smallness of the effective
couplings $\lambda$ and $k$ in the low energy theory \cite{phiNMSSM}.
The high energy theory has no exact global $U(1)_{PQ}$ symmetry,
which emerges as an accidental approximate symmetry of the low energy
effective theory. The solution to the strong CP problem was shown
to be maintained in a particular example with a discrete
$Z_3 \times Z_5$ symmetry \cite{phiNMSSM}.

An open question facing the model is that of the origin of the 
vacuum energy $V(0)$ in Eqs.\ref{vphin2}, \ref{V0}.
We have neither explained the origin of this vacuum energy,
nor explained how it exactly cancels the energy of the explicit
potential at the global minimum (which amounts to solving the
cosmological constant problem.) 
Furthermore, since the vacuum energy
is expected to break supersymmetry, one would expect that in the
more general framework of supergravity that it would lead to a contribution
to the inflaton mass of order the Hubble constant $H\approx V(0)^{1/2}/3M_P$
or $H \sim 1 $ MeV which although small is much larger than the
required inflaton mass $\sim 1 $ eV, and would result in an $\eta $
parameter of order unity (the well known $\eta $ problem.)
One possibility, suggested in \ref{phiNMSSM}, is that the vacuum
energy results from the D-term part of the potential.
But such a D-term inflation scenario \cite{riotto} seems 
rather unlikely to work here both since the height of the potential
in this model is very low compared to the string scale, 
$V(0)^{1/4}\sim 10^8$ GeV, and since the solution to the
cosmological constant problem would require an
even more miraculous cancellation than usual, since the explicit
potential results from F-terms. A better possibility would seem to be
to appeal to the string
no-scale supergravity framework which is supposed to
account for the masslessness of the $\phi$ field, which we identify here
as a moduli-like scalar \cite{noscale}. In such a framework one can
hope to fine-tune the cosmological constant to zero whilst maintaining
a vacuum energy during inflation. Moreover in a certain class of no-scale
model (those in which a Heisenberg symmetry is present)
the inflaton receives no mass of order the Hubble constant
thereby solving the $\eta$ problem \cite{eta3}.
The challenge is to find an explicit string no-scale supergravity
model which does this, and at the same time allows the soft couplings
that we need in our model.

\section*{Acknowledgments}
We would like to thank Leszek Roskowski and all the other
organisers and participants (especially Mark Hindmarsh, David Lyth
Toni Riotto, Jenny Sanderson and many others) for making COSMO-97 such 
a stimulating and enjoyable experience.

\end{document}